\documentclass[a4paper,11pt]{article}
\pdfoutput=1 

\usepackage{jheppub} 

\usepackage[T1]{fontenc} 
\usepackage{slashed}

\newcommand{\bfm}[1]{\mbox{\boldmath$#1$}}

\newcommand{\gsim}{\;\rlap{\lower 3.5 pt \hbox{$\mathchar \sim$}} \raise 1pt
\hbox {$>$}\;}
\newcommand{\lsim}{\;\rlap{\lower 3.5 pt \hbox{$\mathchar \sim$}} \raise 1pt
\hbox {$<$}\;}

\title{\boldmath  A Theory of Giant Vortices}

\preprint{ALBERTA-THY-2-21}

\author[a]{Alexander A. Penin} \author[]{and}
\author[a,b]{Quinten Weller}

\affiliation[a]{Department of Physics, University of Alberta,
Edmonton AB T6G 2E1, Canada
}
\affiliation[b]{Department of Mathematical and Statistical Sciences, University of Alberta,
Edmonton AB T6G 2G1, Canada
}

\emailAdd{penin@ualberta.ca}
\emailAdd{qweller@ualberta.ca}

\abstract{We elaborate a theory of  {\it giant} vortices
\cite{Penin:2020cxj} based on  an asymptotic expansion in
inverse powers of their  winding number $n$. The theory is
applied to the analysis of vortex solutions in the abelian
Higgs (Ginzburg-Landau) model. Specific properties of the giant
vortices for charged and neutral scalar fields as well as
different  integrable limits of the scalar self-coupling are
discussed. Asymptotic results  and the  finite-$n$ corrections
to the vortex solutions are derived in analytic form and the
convergence region of the expansion is determined.}

\begin{document}
\maketitle
\flushbottom
\section{Introduction}
\label{sec::int}

Vortices, string-like topologically nontrivial solutions of
field equations, naturally appear in the theory of
superconductivity \cite{Abrikosov:1956sx}, superfluidity
\cite{Ginzburg:1958},  and QCD confinement
\cite{Nielsen:1973cs}. They play a crucial role in many
physical concepts   from cosmic strings \cite{Hindmarsh:1994re}
and vacuum selection \cite{Penin:1996si} to  mirror symmetry
and dualities of supersymmetric models \cite{Tong:2005un}. {\it
Giant}  vortices carrying large topological charge  are of
particular interest and are observed experimentally in a variety
of quantum condensed matter systems
\cite{Marston:1977,Engels:2003,Cren:2011}. Corresponding
winding numbers  range  from $n=4$ in  mesoscopic
superconductors \cite{Cren:2011} through $n=60$ in
Bose-Einstein condensate of cold atoms \cite{Engels:2003} and
up to  $n=365$ in superfluid $^4$He \cite{Marston:1977}. From a
theoretical perspective it is quite  appealing  to identify
characteristic features and universal properties of vortices in
the limit of large $n$. A solution to this problem is too
subtle for a straightforward numerical analysis and requires
some form of analytic approach.

Though the system of vortex equations is relatively simple, its
exact analytic solution is not available even for critical
coupling  when hidden supersymmetry reduces the order of the
equations \cite{Bogomolny:1975de} and even for the lowest
winding number $n=1$  \cite{deVega:1976xbp}  in contrast to the
apparently more complex case of magnetic monopoles
\cite{Prasad:1975kr}. In  general, finding analytic solutions
of higher topological charge is a very challenging problem  and
only a few such solutions are known in gauge models (see {\it
e.g.} \cite{Witten:1976ck,Prasad:1980hg}). By contrast  the
structure of vortices vastly simplifies in the limit of large
winding number $n\to\infty$
\cite{Bolognesi:2005rj,Bolognesi:2005zr}.  In a recent  letter
\cite{Penin:2020cxj} we have introduced  a framework which
enables a systematic  expansion in inverse powers of $n$ to
obtain the asymptotic form of the axially symmetric giant
vortex  solution. In this framework a topological quantum
number $n$  is  associated with a ratio of dynamical scales and
a systematic expansion in inverse powers of $n$ is then derived
in the spirit of  effective field theory. Below we present a
detailed account of the method and its application to the
Abrikosov-Nielsen-Olesen (charged field) and
Ginzburg-Pitaevskii (neutral field) vortices.

\section{Abrikosov-Nielsen-Olesen vortices}
\label{sec::2}

We consider the standard Lagrangian for the abelian Higgs
(Ginzburg-Landau) model of a scalar field with  abelian charge $e$,
quartic  self-coupling $\lambda$, and vacuum expectation value $\eta$
in two dimensions
\begin{equation}
{L}=-{1\over 4} F^{\mu\nu}F_{\mu\nu}
+\left({D^\mu \phi}\right)^\dagger D_\mu \phi
-{\lambda\over 2}\left(\left|\phi\right|^2-\eta^2\right)^2\,,
\label{eq::Lagrange}
\end{equation}
where $D_\mu=\partial_\mu+ieA_\mu$. Vortices are topologically
non-trivial solutions of the corresponding Euclidean equations
of motion. We  study the  axially symmetric solutions  of
winding number $n$, which in polar coordinates can be written
as follows $\phi(r,\theta)=f(r)e^{in\theta}$, $A_\theta=-n
a(r)/e$, $A_r=0$. For a given winding number the solution
carries $n$ quanta of magnetic flux $\Phi= -\int F_{12}{\rm
d}^2{\bfm r}=2\pi n$. When the winding number grows, the
characteristic size of the vortex has to grow as well  to
accommodate the  increasing magnetic flux. Assuming a roughly
uniform average distribution of the flux inside the vortex for
$\lambda,\hspace*{1pt}\eta={\cal O}(1)$ we can estimate  its radius to be
of order $\sqrt{n}/e$.  At the same time a characteristic
distance of the nonlinear interaction is $1/e$. Thus for large
$n$ we get a scale hierarchy and the expansion in the
corresponding scale ratio is a standard tool  of the effective
field theory approach. Since we deal with the spatially
extended classical solutions it is more convenient to perform
this expansion in coordinate space at the level of the
equations of motion. The analysis becomes particulary simple
for critical  coupling $\lambda=e^2$, which we discuss first.

\subsection{Critical vortices}
\label{sec::2.1}
In this case the vortex dynamics is governed by
the  first-order Bogomolny equations \cite{Bogomolny:1975de}
\begin{equation}
\begin{split}
& \left(D_1+iD_2\right)\phi =0\,, \\
& -F_{12}+e\left(\left|\phi\right|^2-\eta^2\right)=0\,,
\label{eq::BogomolnyphiF}
\end{split}
\end{equation}
and the vortex energy (string tension) is proportional to the
topological charge  $T= -\int L{\rm d}^2{\bfm r}=2\pi n\eta^2$.
It is convenient to introduce  the rescaled dimensionless
quantities $e\eta r\to r$, $f/\eta\to f$,
$\lambda/e^2\to\lambda$, so that in the new variables $e=\eta=1$
and critical coupling corresponds to $\lambda=1$. Then the
Bogomolny equations in terms of the functions $a(r)$ and $f(r)$
take the following form
\begin{equation}
\begin{split}
& {df\over dr}-{n\over r}(1-a)f  = 0 \,, \\
& {da\over dr}+{r\over n}(f^2-1) = 0\,,
\label{eq::Bogomolnyfa}
\end{split}
\end{equation}
with the boundary conditions $f(0)=a(0)=0$ and
$f(\infty)=a(\infty)=1$. For large $n$ the field dynamics is
essentially different in three regions: the core,   the
boundary layer, and the tail of the vortex. Below we discuss
the specifics of the dynamics and its description in each
region.

\noindent
{\em The vortex core.} For small $r$ the solution of the field
equations gives $f(r)\propto r^n$. This function is
exponentially suppressed at large $n$ for all $r$ smaller than
a critical value, which  can be associated with the core
boundary. For such $r$ the contribution of $f(r)$ can be
neglected in the equation for $a(r)$ and we get $a(r)\sim
r^2/r_n^2$ with $r_n=\sqrt{2n}$, which in turn can be used in
the equation for $f(r)$. Thus in the core the dynamics is
described by linearized equations in the background field
\begin{equation}
\begin{split}
& {df\over dr}-{n\over r}\left(1-{r^2\over r_n^2}\right)f  = 0 \,, \\
& {da\over dr}-{r\over n} = 0\,.
\label{eq::coreeq}
\end{split}
\end{equation}
Their  solutions read
\begin{equation}
\begin{split}
&f(r)=F\exp\left[{n\over 2}\left(\ln\left({r^2\over r^2_n}\right)
- {r^2\over r_n^2}+1\right)\right],\\
&a(r)={r^2\over r_n^2}\,,
\label{eq::coresol}
\end{split}
\end{equation}
where the integration constant $F$ in the first line is
determined by matching conditions explained below. For
$r_n-r={\cal O}(1)$ we have  $n(1-a(r))/r={\cal O}(1)$ and the
equation for $f(r)$ becomes independent of $n$. Hence the
approximation Eq.~(\ref{eq::coreeq}) is not applicable anymore,
the nonlinear effects become crucial, and  we enter the
boundary layer.  Note that the magnetic flux and energy density
for Eq.~(\ref{eq::coresol}) are approximately $1$ and $\eta^2$,
respectively, so that  the core accommodates essentially all
the vortex flux and energy and  we can identify $r_n$ with the
vortex radius.

\noindent
{\em The  boundary layer.} In this region the field dynamics is
ultimately nonlinear. However, it crucially simplifies  for large $n$.
To see this we introduce a new radial coordinate $x=r-r_n$ so that in
the boundary layer $x = {\cal O}(1)$ and the expansion in $x/r_n$
converts into an expansion in $1/\sqrt{n}$. In the leading order in
$x$ Eq.~(\ref{eq::Bogomolnyfa}) reduces to  a system of
$n$-independent field equations with constant coefficients
\begin{equation}
\begin{split}
& w'+\gamma= 0 \,, \\
& \gamma' -1+e^{2w}=0\,,
\label{eq::boundeq}
\end{split}
\end{equation}
where  $w(x)=\ln f(r_n+x)$,
$\gamma(x)={n}\left(a(r_n+x)-1\right)/r_n$, and prime stands for a
derivative in $x$.  The system can be resolved for $w$, which results
in  a second-order equation
\begin{equation}
\begin{split}
& w''+1-e^{2w}= 0\,.
\label{eq::weq}
\end{split}
\end{equation}

\begin{figure}[t]
\begin{center}
\includegraphics[width=8cm]{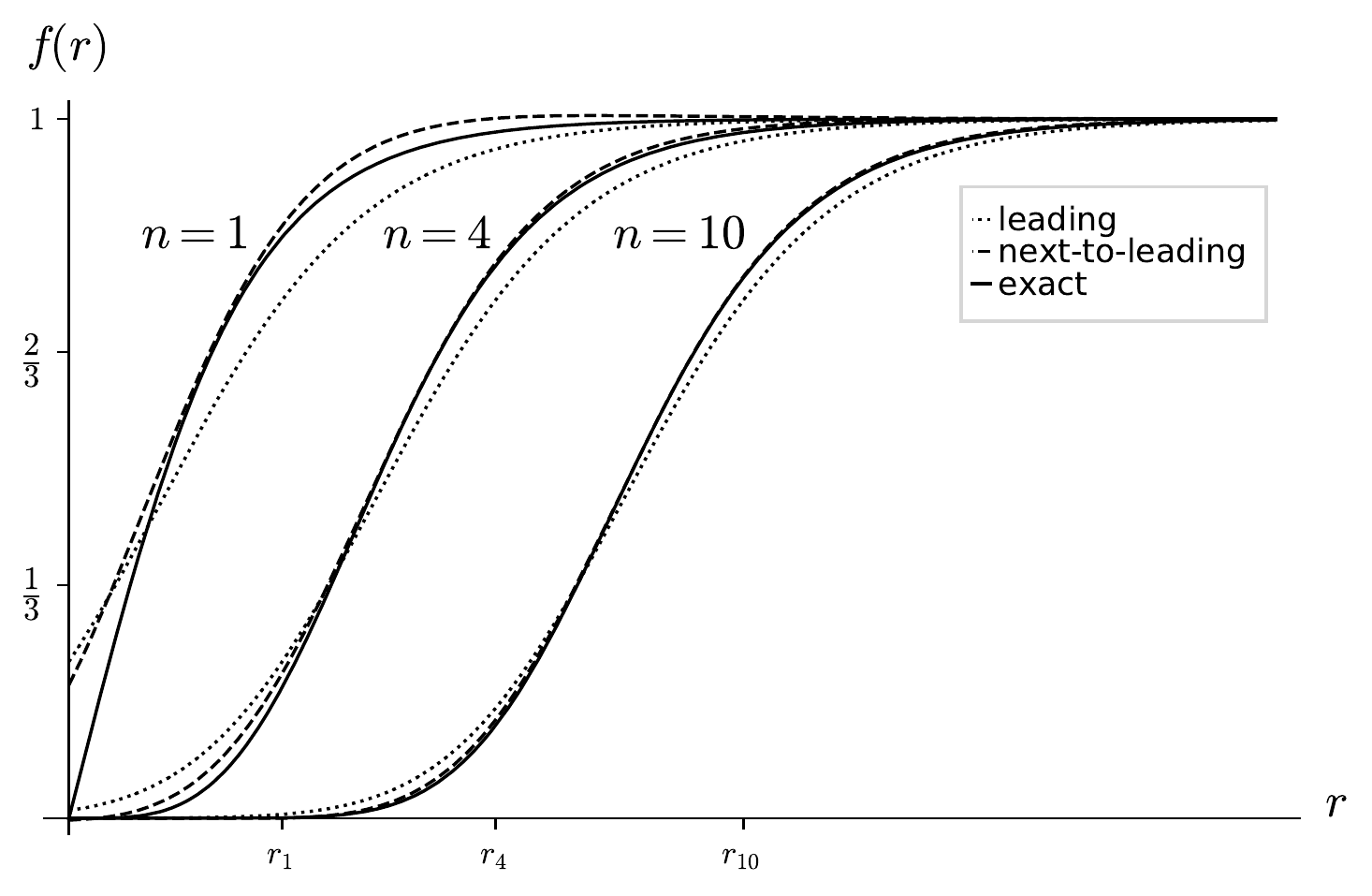}
\end{center}
\caption{\label{fig::1}   The  numerical  solution of the exact
critical vortex equations for the scalar field $f(r)$ (solid lines),
the leading asymptotic boundary layer solution (dotted lines), and
the next-to-leading approximation  (dashed lines) for different
winding numbers $n$.}
\end{figure}

\noindent
This equation  has a first integral $I=w'^2-e^{2 w}+2w$ with $I=-1$
corresponding to the boundary condition $w(\infty)=0$. Thus
Eq.~(\ref{eq::boundeq}) can be solved in quadratures with the result
\begin{equation}
\begin{split}
& \int^{w(x)}_{w_0} {{\rm d}w \over (e^{2 w}-2w-1)^{1/2}}=x  \,,\\
& \gamma(x) =  -(e^{2 w(x)}-2w(x)-1)^{1\over 2}\,,
\label{eq::boundsol}
\end{split}
\end{equation}
where $w_0=w(0)$  is the second  integration constant.  It is
determined by the boundary condition $w'(x)\sim -x$ at $x\to-\infty$,
which ensures that Eq.~(\ref{eq::boundsol}) can be matched to the
core solution. This gives a new transcendental constant
\begin{equation}
w_0=-0.2997174398\ldots\,,
\label{eq::wellerconst}
\end{equation}
which determines a unique asymptotic solution in the boundary
layer. It has a Taylor expansion  $w(x)=\sum_{m=0}^\infty
w_mx^m$ where $w_1=(e^{2w_0}-2w_0-1)^{1/2}$  and the higher
order coefficients can be obtained recursively
\begin{equation}
w_2= {e^{2w_0}-1\over 2} \,,
\quad w_3= {w_1\over 3}e^{2w_0} \,,
\quad w_4= {w_1^2+w_2\over 6}e^{2w_0} \,,
\quad \ldots \,.
\label{eq::taylorw}
\end{equation}
The asymptotic behavior of the function at $x\to\infty$ reads
\begin{equation}
\begin{split}
& w(-x) \sim -{x^2\over 2}-{1\over 2}\,,\\
&w(x) \sim w_\infty e^{-\sqrt{2}x}  \,,
\label{eq::wasym}
\end{split}
\end{equation}
\noindent
where the constant
\begin{equation}
w_\infty=w_0\exp\left[\int_{w_0}^0\left({\sqrt{2}\over (e^{2w}
-2w-1)^{1/2}}+{1\over w}\right){\rm d}w\right]=-0.331186\ldots\,
\label{eq::winfty}
\end{equation}
is computed in Appendix~\ref{sec::appA}. By  matching
Eq.~(\ref{eq::wasym})  to the core solution,
Eq.~(\ref{eq::coresol}),  in the region $1\ll r_n-r \ll r_n$
where both approximations are valid we get  $F=1/\sqrt{e}$.

\begin{figure}[t]
\begin{center}
\includegraphics[width=8cm]{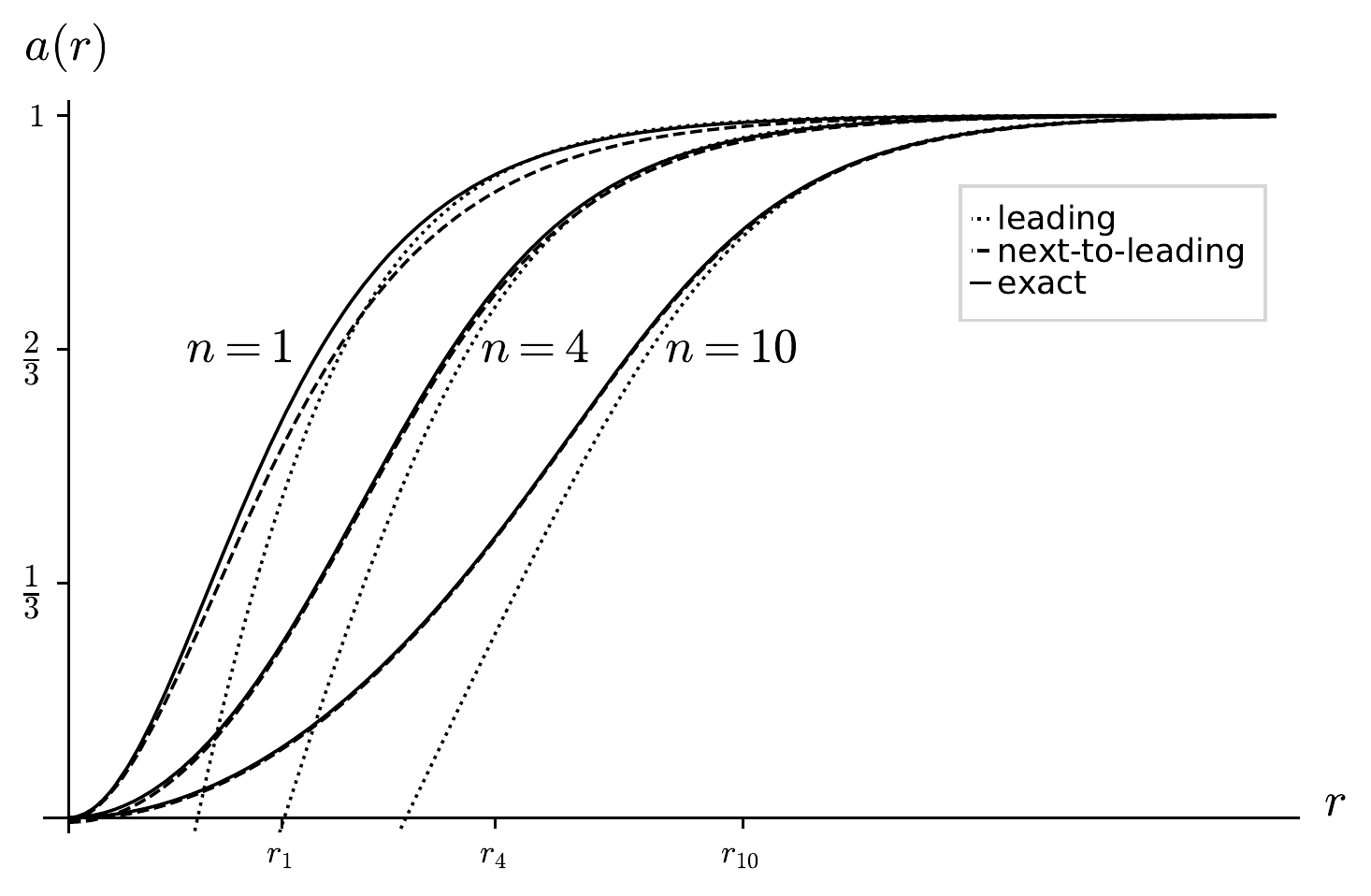}
\end{center}
\caption{\label{fig::3}  Same as Fig.~\ref{fig::1} but for
the gauge field $a(r)$.}
\end{figure}

\noindent
{\em The vortex tail.} For $(r-r_n)/r_n={\cal O}(1)$  the boundary
layer approximation breaks down and the coordinate dependence of the
field equation coefficients should be restored. However, the deviation
of the fields from the vacuum configuration is now exponentially
small so the field  equations linearize. The solution of the
linearized theory is well known and reads
\begin{equation}
\begin{split}
& f(r) \sim 1+ {\nu\over 2\pi}K_0(\sqrt{2}r)+\ldots\,, \\
&a(r) \sim 1+ {\mu\over 2\pi}{\sqrt{2}r}K_1(\sqrt{2}r)+\ldots\,,
\label{eq::tailsol}
\end{split}
\end{equation}
where $K_m(z)$ is the $m$th modified Bessel function. It describes
the field of a point-like source of scalar charge $\nu$ and magnetic
dipole moment $\mu$ with $\nu=\mu$ for critical coupling.
Eqs.~(\ref{eq::boundsol}) and (\ref{eq::tailsol}) should coincide in
the second matching region $1\ll r-r_n \ll r_n$, which yields
\begin{equation}
\nu=4w_\infty \sqrt{\pi}e^{2\sqrt{n}+\ln(n)/4}\,.
\label{eq::charge}
\end{equation}

\noindent
{\em Next-to-leading order solution.} For a finite winding
number  the above asymptotic formulae have a finite accuracy
which may deteriorate as $n$ decreases.  To get control over
the accuracy  and convergence of the large-$n$ expansion we
compute the ${\cal O}(1/\sqrt{n})$ correction to the asymptotic
result.  Let us consider first the boundary layer. Writing the
corrections to the asymptotic solutions   $w$ and $\gamma$ as
$\delta w/\sqrt{2n}$ and $\delta \gamma/\sqrt{2n}$,
respectively, and expanding Eq.~(\ref{eq::Bogomolnyfa}) through
${\cal O}(x/r_n)$ we get the following field equations
\begin{equation}
\begin{split}
&\delta w''-2e^{2w}\delta w=-w'\,,\\
&\delta \gamma(x) = -{x}w'(x)-\delta w'(x)\,,
\label{eq::nloeq}
\end{split}
\end{equation}
{\it i.e.} $\delta\gamma$ is completely determined by the
solution for $\delta w$. The latter obeys the boundary
conditions $\delta w(x)\sim x^3/6+{\cal O}(1)$ at
$x\to-\infty$, $\delta w(\infty)=0$, and can be found in a
straightforward  though not completely obvious way. By equating
the variation of the first integral $I$ to zero and changing
the  variable from $dx$ to $dw=w'dx$ we get the following
first-order equation for the homogeneous part of the solution
$\delta w^h$
\begin{eqnarray}
{d\over dw}\left[\left(e^{2w}-2w-1\right)\frac{d\,\delta w^h}{dw}
-\left(e^{2w}-1\right)\delta w^h\right]
=0\,.
\label{eq::nloeq2}
\end{eqnarray}

\begin{figure}[t]
\begin{center}
\includegraphics[width=8cm]{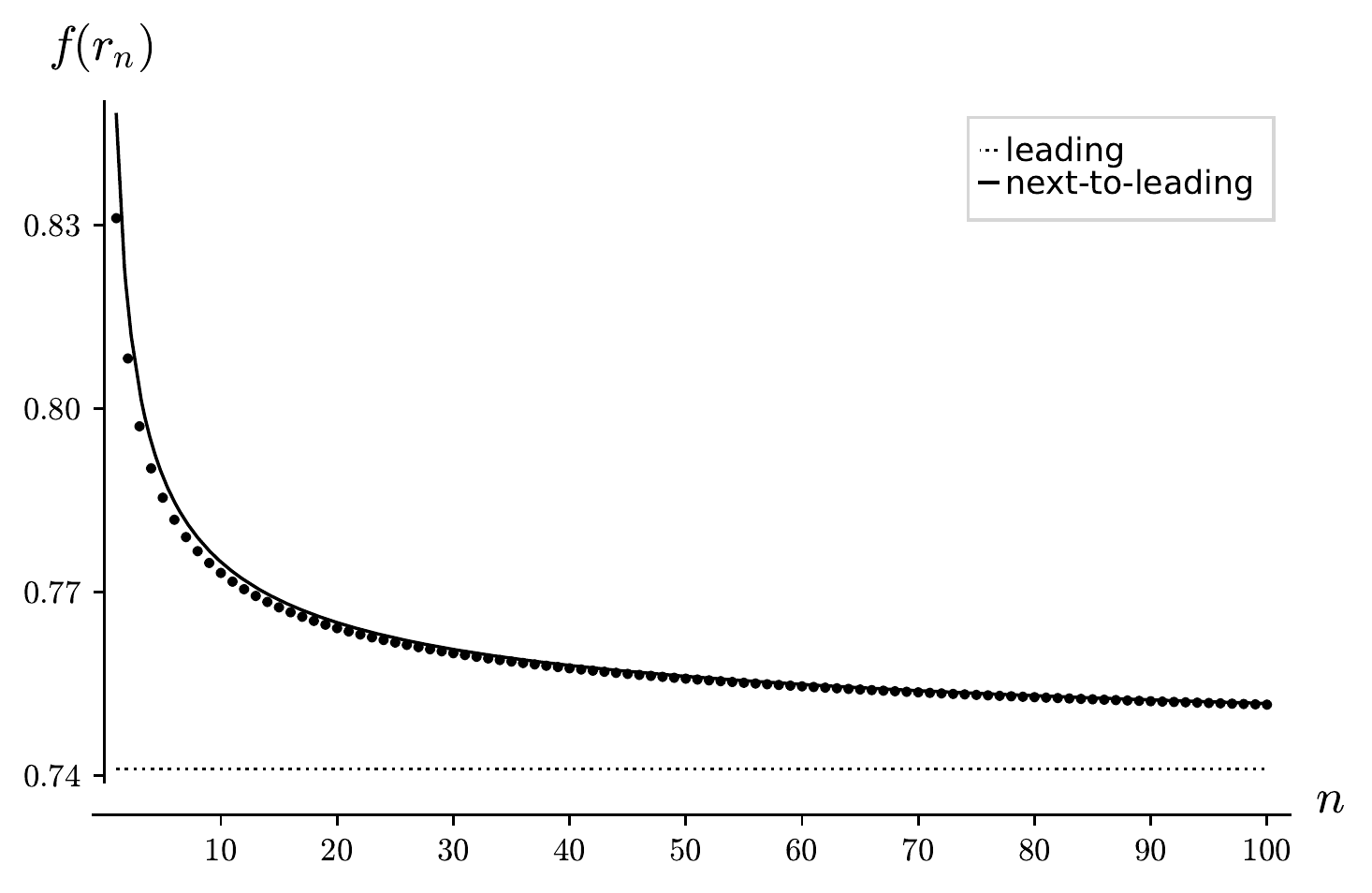}
\end{center}
\caption{\label{fig::2}   The  values $f(r_n)$ obtained from the
numerical solution of the exact critical vortex equations (black
dots), the asymptotic value $f(r_n)=e^{w_0}$ (dotted line), and the
next-to-leading approximation (solid line) as functions of the
winding number $n$.}
\end{figure}

\noindent
One solution of this equation is given by the
translational zero mode $w'(x)$ and we take
$w'(x)\int_{-\infty}^{x}{\rm d}y/w'^2(y)$  as the second
solution so that the pair  has unit Wronskian. This results in
the following solution of Eq.~(\ref{eq::nloeq}) which satisfies
the boundary condition at $x\to\infty$
\begin{equation}
\delta w(x)=w'(x)\int_{x_0}^{x}w'^2(y)
\int_{-\infty}^{y}\frac{1}{w'^2(z)}{\rm d}z{\rm d}y
+w'(x)\int_{-\infty}^{x}\frac{{\rm d}z}{w'^2(z)}
\int_{x}^{\infty}w'^2(y){\rm d}y\,,
\label{eq::nlogensol}
\end{equation}
where $x_0$ is an integration constant  to be determined. A
variation of $x_0$ changes Eq.~(\ref{eq::nlogensol}) by a term
proportional to $w'(x)$ and therefore does not affect its
behavior at $x\to\infty$. For $x_0\to\infty$ the two integrals
in Eq.~(\ref{eq::nlogensol}) can be combined into
\begin{equation}
-w'(x)\int_{x}^\infty \int_x^{y}\frac{w'^2(y)}
{w'^2(z)}{\rm d}z{\rm d}y
=-w'(x)\int_{x}^\infty \int_z^{\infty}
\frac{w'^2(y)}{w'^2(z)}{\rm d}y{\rm d}z\,.
\label{eq::nloint1}
\end{equation}
The last integral up to a term proportional to $w'(x)$ is equal
to
\begin{equation}
w'(x)\int_{0}^x\int_z^{\infty}\frac{w'^2(y)}
{w'^2(z)}{\rm d}y{\rm d}z\,.
\label{eq::nloint2}
\end{equation}
Thus the next-to-leading contribution can be written as follows
\begin{equation}
\delta w(x)=w'(x)\left(C+\int_{0}^{x}\int_{z}^{\infty}
{w'^2(y)\over w'^2(z)}{\rm d}y{\rm d}z\right)\,,
\label{eq::delwres}
\end{equation}
where the integration constant
\begin{equation}
C=\int_{-\infty}^{0} \left( {z\over 3} + \int^{\infty}_{z}
{w'^2(y)\over w'^2(z)} {\rm d}y \right) dz
=0.529935\ldots\,
\label{eq::Cres}
\end{equation}
is adjusted to satisfy the boundary condition at $x\to-\infty$
(see Appendix~\ref{sec::appA}). Eq.~(\ref{eq::delwres})
determines the next-to-leading approximation in the boundary
layer. In the core and tail regions the neglected nonlinear
terms are exponentially suppressed and the corrections to the
asymptotic result enter only through the  matching to the
boundary layer solution. To perform the matching we need the
asymptotic behavior of Eq.~(\ref{eq::delwres}) at
$|x|\to\infty$ which is computed in Appendix~\ref{sec::appA}.
For $x\to\infty$ it reads
\begin{equation}
\begin{split}
& \delta w(-x) \sim -{x^3\over 6}+B\,,\\
&\delta w(x) \sim -w_{\infty}\sqrt{2}{e}^{-\sqrt{2}x}
\left(D+\frac{x}{2\sqrt{2}}\right) \,,
\label{eq::delwasym}
\end{split}
\end{equation}
where
\begin{equation}
B=\int_{-\infty}^{0}\left[{\left(e^{2 w}-2w-1\right)^{1\over 2}}
-\left(-{2w}\right)^{1\over 2}
\left(1+{1\over 4w}\right)\right]{\rm d}w=0.274111\ldots \,,
\label{eq::Bres}
\end{equation}
and
\begin{equation}
D=C+\int_{0}^{\infty}\left(-\frac{1}{2\sqrt{2}}
+\int_{z}^{\infty}\frac{w'^2(y)}{w'^2(z)}{\rm d}y\right){\rm d}z
=0.440087\ldots\,.
\label{eq::Dres}
\end{equation}
The first line of Eq.~(\ref{eq::delwasym}) in the matching
region $0\ll r_n-r \ll r_n$ determines the correction to the
core solution integration constant
\begin{equation}
F=e^{-{1\over 2}+{B\over \sqrt{2n}}}\,.
\label{eq::Fnlo}
\end{equation}
Analogously  the matching of the second line of
Eq.~(\ref{eq::delwasym}) to the tail solution  in the  region
$1\ll r-r_n \ll r_n$ determines the correction to the scalar
charge $\nu$, which we write as $\delta\nu/\sqrt{n}$. Expanding
Eq.~(\ref{eq::tailsol}) at large $r$ and keeping the subleading
term we get
\begin{equation}
f(r)\sim 1+\frac{\nu+\delta\nu/\sqrt{n}}{2\pi}
\left(\frac{\pi}{2\sqrt{2}r}\right)^{1\over 2}
e^{-\sqrt{2}r}\left(1-\frac{1}{8\sqrt{2}r}
+{\cal O}(1/r^2)\right)\,.
\label{eq::ftailasymnlo}
\end{equation}
Further expansion of Eq.~(\ref{eq::ftailasymnlo}) with
$r=r_n+x$ in $x/r_n$ and matching to  the second line of
Eq.~(\ref{eq::delwasym}) gives
\begin{equation}
{\delta\nu\over \nu}={1\over 16}-D=-0.377587\ldots\,,
\label{eq::delnu}
\end{equation}
which completes the next-to-leading approximation of the critical
vortex solution.

\begin{figure}[t]
\begin{center}
\includegraphics[width=8cm]{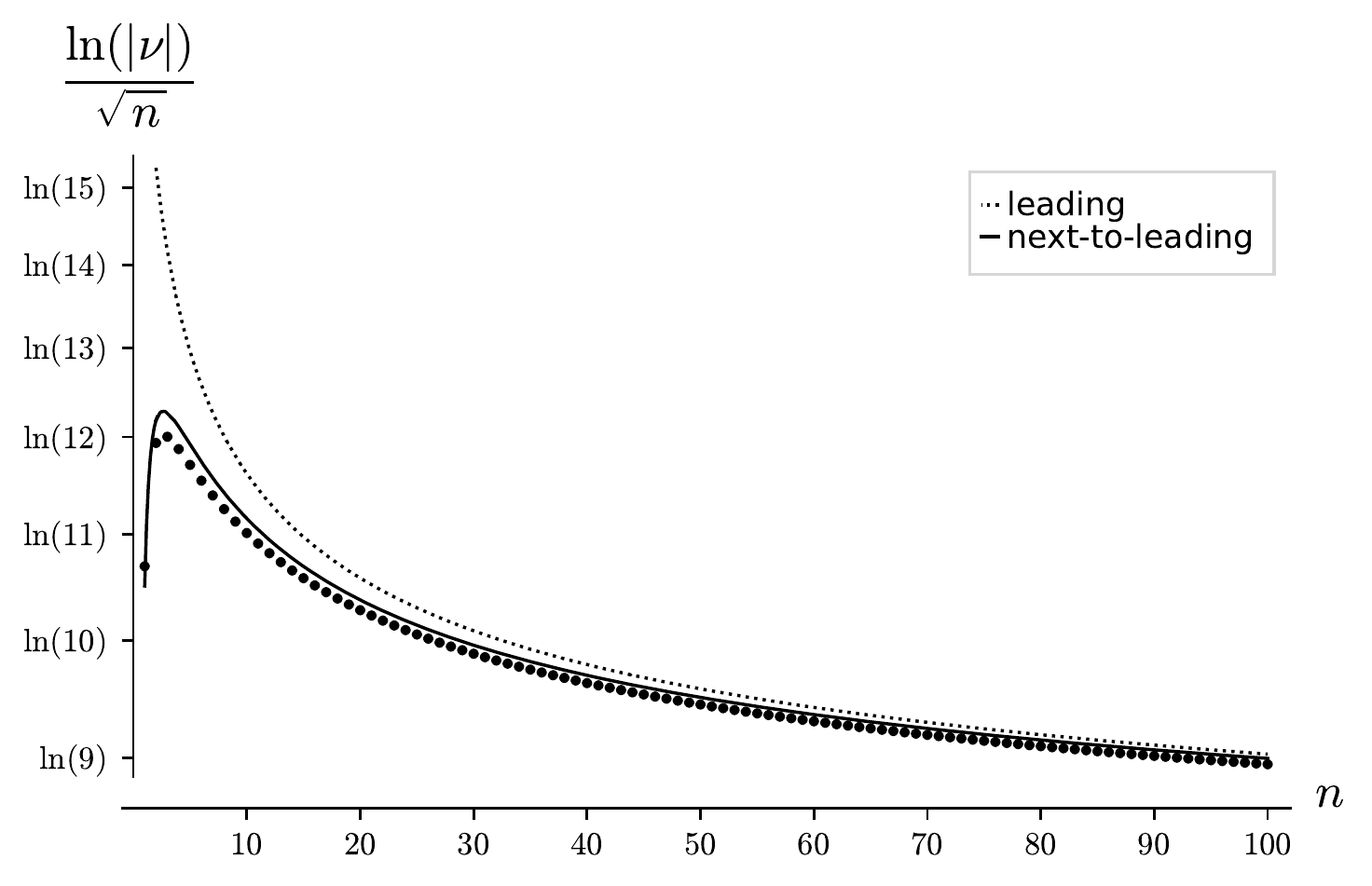}
\end{center}
\caption{\label{fig::4} Scalar charge/magnetic dipole moment  $\nu$
of a critical vortex obtained from the numerical  solution of the
exact vortex equations (black dots), the asymptotic result
Eq.~(\ref{eq::charge})  (dotted line), and the next-to-leading
approximation (solid line) as functions of the winding number $n$.}
\end{figure}

We can now scrutinise the accuracy and convergence of the
large-$n$ expansion by comparing the leading and
next-to-leading approximations to  numerical solutions of the
exact field equations for different winding numbers. The
results of the analysis for the critical vortices are presented
in Figs.~\ref{fig::1}-\ref{fig::4}. In Fig.~\ref{fig::1} the
leading and next-to-leading boundary layer solutions for
the scalar field $f(r)$ are plotted against the exact
(numerical) solutions  with  $n=1,\,4,\,10$. Similar plots for
the gauge field $a(r)$ are given in Fig.~\ref{fig::3}. In
Figs.~\ref{fig::2} and \ref{fig::4} the exact numerical values
of $f(r_n)$ and $\nu$, the natural characteristics of the
vortex solution, are plotted against the leading and
next-to-leading results for $n=1,\ldots,100$. The expansion
reveals an impressive convergence and the next-to-leading
approximation works reasonably well even for $n=1$.

\subsection{Noncritical vortices}
\label{sec::2.2}
\begin{figure}[t]
\begin{center}
\includegraphics[width=8cm]{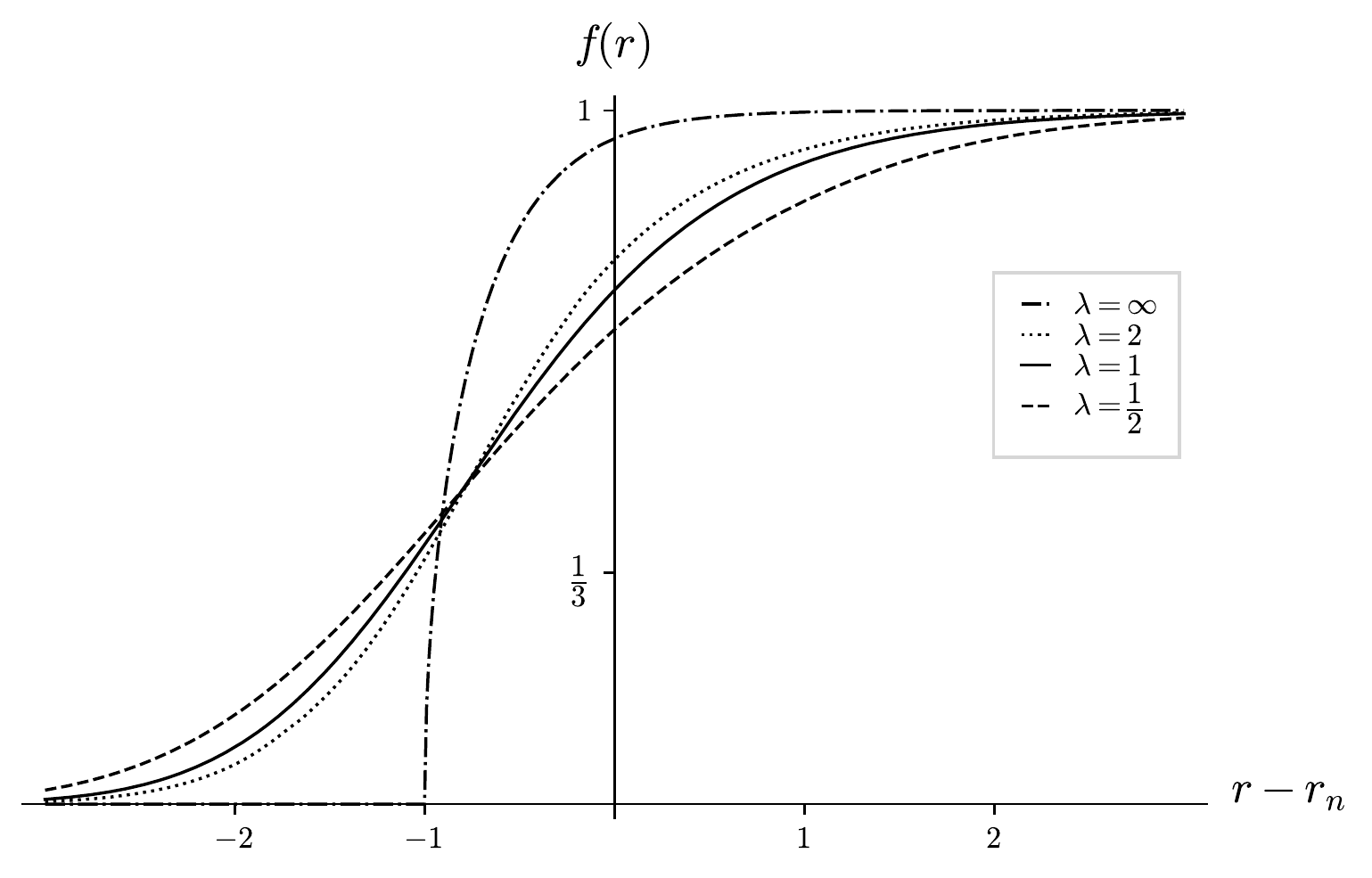}
\end{center}
\caption{\label{fig::5}   The  asymptotic profiles of the
scalar field $f(r)$ obtained by  numerical  solution of the
boundary layer equations for different values of scalar
self-coupling and the analytic result
Eq.~(\ref{eq::largelamsol}) for $\lambda\to\infty$ as functions
of $r-r_n$. For $\lambda\to 0$ the scalar field vanishes
throughout the boundary layer as $\lambda^{1/4}$.}
\end{figure}

\begin{figure}[t]
\begin{center}
\includegraphics[width=8cm]{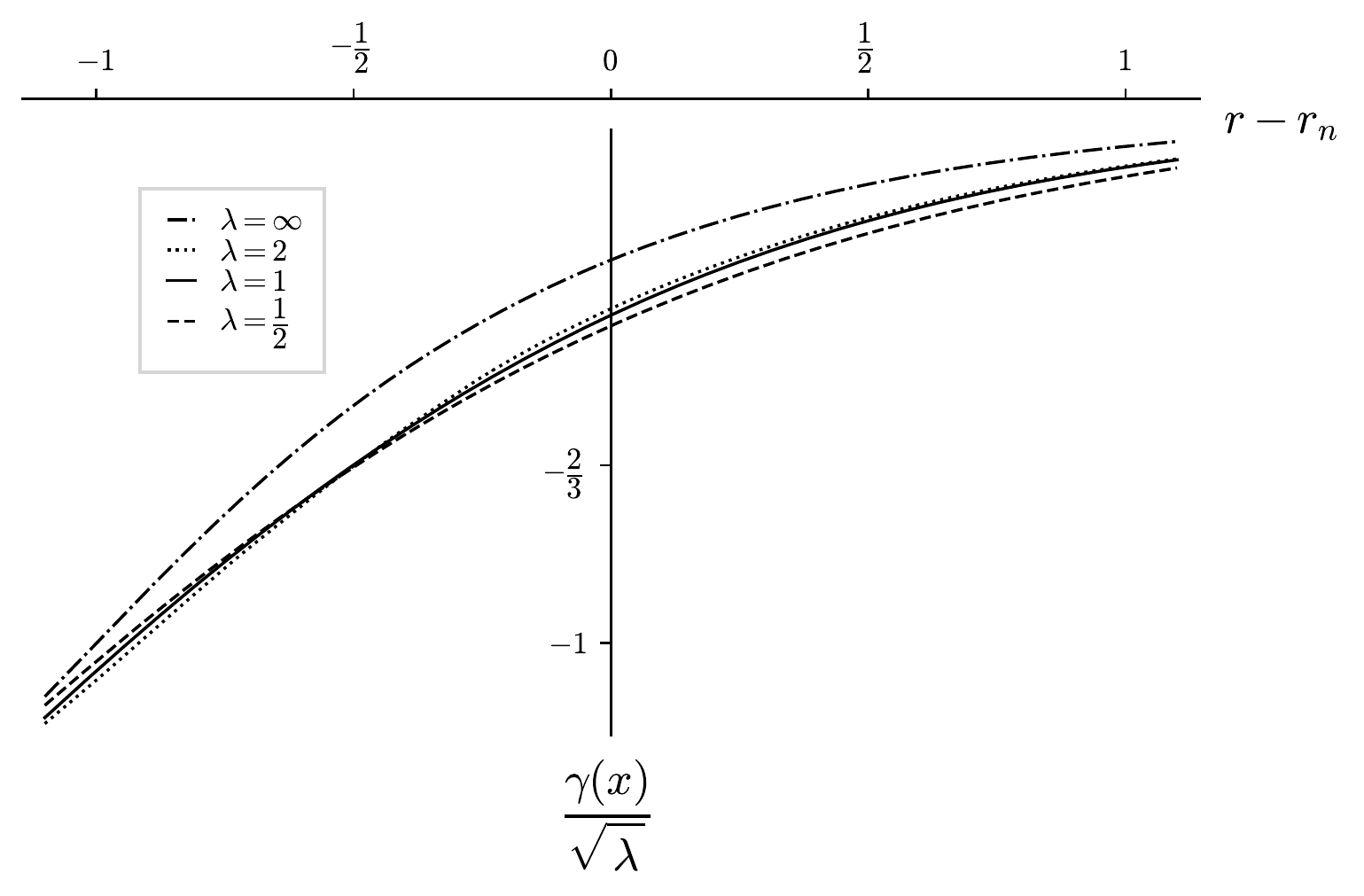}
\end{center}
\caption{\label{fig::6}  Same as Fig.~\ref{fig::5} but for
the normalized gauge field $\gamma(x)/\sqrt{\lambda}$.}
\end{figure}

For noncritical scalar self-coupling $\lambda\ne 1$ the order
of the field equations  cannot be reduced and they read
\begin{equation}
\begin{split}
&{1\over r}{d\over dr}\left(r{df\over dr}\right)-
 \left[\lambda(f^2-1)+{n^2\over r^2}(1-a)^2\right]f  = 0 \,,\\
& r{d\over dr}\left({1\over r}{da\over dr}\right)+{2}(1-a)f^2 = 0\,.
\label{eq::fieldeqfa}
\end{split}
\end{equation}
Nevertheless, the general structure of the solution  is quite
similar to the critical case. Inside the core the contribution
of the scalar potential to Eq.~(\ref{eq::fieldeqfa}) is
suppressed by $r^2/n^2$. Hence the core dynamics is not
sensitive to  $\lambda$ and the core solution is given by
Eq.~(\ref{eq::coresol}) up to the value of the integration
constants which do depend on $\lambda$ through the matching to
the non-linear boundary layer solution. In particular the
vortex size $r_n$ is  determined by the region where the two
terms in the square brackets of Eq.~(\ref{eq::fieldeqfa})
become  comparable and the core approximation breaks down,
which gives the leading order result
$r_n=\sqrt{2n}/\lambda^{1/4}$. Note that the approximately
constant energy density in the core is now $\lambda\eta^2$ so
that the  total vortex energy in the large-$n$ limit  is
$T=2\pi\sqrt{\lambda}n\eta^2$. In the tail solution,
Eq.~(\ref{eq::tailsol}), the argument of $K_0$ gets an
additional factor of $\sqrt{\lambda}$ to account for the
variation of the scalar field mass, while the scalar charge and
the magnetic dipole moment are not equal anymore and have
different leading behavior  at $n\to \infty$
\begin{equation}
\begin{split}
&|\nu|\sim e^{2\sqrt{n}\,\lambda^{1/4}+\ldots} \,,\qquad
|\mu|\sim e^{2\sqrt{n}/\lambda^{1/4}+\ldots}\,.
\label{eq::chargelam}
\end{split}
\end{equation}
More accurately these parameters as well as the normalization
of the scalar field in the core solution are determined by
matching to the boundary layer solution. In the boundary layer
by expanding $r=r_n+x$ in $x/r_n$ we get a system of
$n$-independent equations with constant coefficients
\begin{equation}
\begin{split}
&{f''}-
 \left[{\lambda}\left({f^2}-1\right)+{\gamma^2}\right]f  = 0 \,,\\
& \gamma''-{2}\gamma{f^2} = 0\,,
\label{eq::domwall}
\end{split}
\end{equation}
and the boundary conditions $\gamma(x)\sim \sqrt{\lambda}x$ at
$x\to-\infty$, $\gamma(\infty)=f(-\infty)=0$, and
$f(\infty)=1$. These equations describe  a domain wall
separating the regions of broken and unbroken symmetry phases
in the one-dimensional effective field theory. For $\lambda=1$
the proper  solution is given by Eq.~(\ref{eq::boundsol}), and
for any given $\lambda\ne 1$ it can be found numerically. The
asymptotic profiles of the boundary layer solution  with
$\lambda=1/2,\, 1,\, 2$ are plotted in Figs.~\ref{fig::5}
and~\ref{fig::6}. In general for $\lambda\ne 1$ the analytic
result for the  boundary layer solution is  out of  reach.
However, in a number of special limits the boundary layer
dynamics vastly simplifies. Below we consider these limits
which gives crucial insight into the convergence and accuracy
of the large-$n$ expansion for the noncritical case.

\noindent
{\em Near-critical coupling.}
For small deviation  $\delta\lambda\equiv \lambda-1$ from the
critical value an expansion in $\delta\lambda$ about
the critical vortex solution can be employed. In particular the
corrections in $\delta\lambda$ can be retained only in the
leading asymptotic result while the critical coupling result of
the previous section can be used for the  ${\cal
O}(1/\sqrt{n})$ terms. Note that the vortex energy is not
proportional to the topological charge anymore and gets a
${\cal O}(\delta\lambda/\sqrt{n})$ contribution which can be
easily evaluated. Indeed we can consider the
$-\delta\lambda(|\phi|^2-\eta^2)^2/2$ term as  a perturbation
to the critical coupling Lagrangian and get the correction to
the vortex energy by evaluating the perturbation on the
critical vortex solution. This gives
\begin{eqnarray}
\delta T&=&{\eta^2\delta\lambda\over 2}
\int \left(1-f^2(r)\right)^2{\rm d}^2{\bfm r}
\nonumber\\
&=&{\eta^2\delta\lambda\over 2}\left[\int
\left(1-f^2(r)\right){\rm d}^2{\bfm r}
-\int \left(1-f^2(r)\right)f^2(r)
{\rm d}^2{\bfm r}\right]\,.
\label{eq::delT}
\end{eqnarray}
Here the first integral in the brackets is the magnetic flux
$2\pi n$ while the second integral is saturated in the boundary
layer with an exponential accuracy. Thus in the latter  we can
approximate the volume element as follows   ${\rm d}^2{\bfm
r}\to 2\pi \sqrt{2n} {\rm d}x= 2\pi \sqrt{2n} {\rm d}w/w'$ and
transform it into $-2\pi \sqrt{n}\sigma$, where
\begin{equation}
\sigma =\int_{-\infty}^0 {\sqrt{2}(1-e^{2w})e^{2w}
\over(e^{2 w}-2w-1)^{1/2}}{\rm d}w=0.775304\ldots\,.
\label{eq::sigma}
\end{equation}
The correction to the critical vortex energy then reads
\begin{equation}
\delta T=
\pi n\eta^2\delta\lambda\left(1-
{\sigma\over \sqrt{n}}+{\cal O}(1/n)\right)\,,
\label{eq::delTres}
\end{equation}
where the first term originates from the expansion of the
asymptotic result $2\pi\sqrt{\lambda}n\eta^2$ in
$\delta\lambda$ and the second term represents the correction
to the boundary layer  energy. Thus the energy of a
near-critical giant vortex to the first order in
$\delta\lambda$ can be written as follows
\begin{equation}
T=2\pi\sqrt{\lambda}n\eta^2\left(1
-{\delta\lambda\,\sigma\over 2\sqrt{n}}
+{\cal O}(\delta\lambda/n)\right)\,.
\label{eq::Tlam}
\end{equation}

\noindent
{\em Large scalar self-coupling.}
In the limit $\lambda\to\infty$, after a field rescaling
$\gamma_\lambda(x)=\gamma(x)/\sqrt{\lambda}$ we may neglect the
derivative term in the first line of Eq.~(\ref{eq::domwall}),
which becomes algebraic   and can be solved for the function
$f(r)$. The field equations then become
\begin{equation}
\begin{split}
&f^2-1+{\gamma_\lambda^2}  = 0 \,,\\
& \gamma_\lambda''-2\gamma_\lambda+2\gamma_\lambda^3 = 0\,.
\label{eq::largelameq}
\end{split}
\end{equation}
The second line of Eq.~(\ref{eq::largelameq}) has a first
integral
${\gamma_\lambda'}^2-2\gamma_\lambda^2+\gamma_\lambda^4=0$
corresponding to the boundary value $\gamma_\lambda(\infty)=0$.
In contrast to Eq.~(\ref{eq::domwall}) it can be integrated in
terms of elementary functions
$\gamma_\lambda(x)=-\sqrt{2}\,{\rm
sech}\left(\sqrt{2}(x-x_0)\right)$, where  $x_0$ is the
remaining integration constant  formally determined by the
boundary conditions at $x\to-\infty$.  However, for large scalar
self-coupling these boundary conditions are rather peculiar
since the matching region shrinks to a point $x=0$. Indeed, in
the matching region the scalar field is exponentially
suppressed $f(r)= {\cal O}(e^{-\sqrt{\lambda} x^2/2})$ and
at $\lambda\to\infty$ it vanishes for all $x<0$ and  at $x=0$
its derivative is infinite. At the same time for negative $x$
the gauge field is given by the core solution and has a
continuous first derivative  at $x=0$. These conditions are
satisfied for $x_0=-{\rm arcsinh}\left(1\right)/\sqrt{2}$. Thus
the boundary layer solution for $x\ge 0$ reads
\begin{equation}
\begin{split}
&f(r_n+\delta r_n+x) =
\left(1-{\gamma^2(x)\over\lambda}\right)^{1\over 2} \,,\\
& \gamma(x) = -{\sqrt{2\lambda}}\,{\rm sech}
\left(\sqrt{2}(x-x_0)\right)\,,
\label{eq::largelamsol}
\end{split}
\end{equation}
while   $f(r)=0$,  $a(r)=r^2/r_n^2$ inside the core for $r \le
r_n+\delta r_n$, where $\delta r_n =-1$. Note that now
$r=r_n+\delta r_n+x$, {\it i.e.}  the core radius which defines
the origin of the $x$ coordinate gets a ${\cal
O}(\lambda^{1/4}/\sqrt{n})$ correction. This correction makes
the winding number of the above gauge field configuration an
integer. We can now compute the finite-$n$ correction to the
vortex energy. As in the case of near-critical coupling it
comes from two sources. Due to the  variation  of the radius
the core energy changes by $-1$ in the units of
$2\pi\lambda^{3/4}\sqrt{2n}\eta^2$. The contribution of the
boundary layer in the same units is ${2\sqrt{2}-1\over 3}$. In
total this gives
\begin{equation}
T=2\pi\sqrt{\lambda}n\eta^2\left(1
-{4(\sqrt{2}-1)\over 3}{\lambda^{1/4}\over \sqrt{n}}
+{\cal O}(1/\sqrt{n})\right)\,.
\label{eq::Tlamlarge}
\end{equation}
Eq.~(\ref{eq::Tlamlarge}) is consistent with the known negative
surface tension on the phase interface in extreme type-II
superconductors \cite{Thomas}. The boundary layer solution,
Eq.~(\ref{eq::largelamsol}), is subject to enhanced ${\cal
O}(\lambda^{1/4}/\sqrt{n})$ corrections which can be obtained
in analytic form by the method of  Sec.~\ref{sec::2.1}. It is
convenient to  write the corrections to $f(r_n+\delta r_n+x)$
and $\gamma_\lambda(x)$ as ${\lambda^{1/4}\over
\sqrt{2n}}\delta\hspace*{-1pt}f(x)$ and ${\lambda^{1/4}\over
\sqrt{2n}}\delta \gamma_\lambda(y)$, where $y=x-x_0$. Then for
the gauge field  we get (see Appendix~\ref{sec::appB})
\begin{eqnarray}
\delta\gamma_\lambda(y)&=& {e^{-\sqrt{2}y}\over 6}
\left(1+ e^{2\sqrt{2}y}\right)^{-2}
\left[1+\left(4 + 10\sqrt{2}
- 18\sqrt{2} (y+x_0)\right)e^{2\sqrt{2}y}
\right.\nonumber\\
&+&\left.\left(-13 + 14\sqrt{2}
- 6\sqrt{2}(y+x_0)\right)e^{4\sqrt{2}y}\right]\,,
\label{eq::delgres}
\end{eqnarray}
and the result for the scalar field is obtained by substituting
Eq.~(\ref{eq::delgres}) into the first line of  Eq.~(\ref{eq::delgeq}).

\noindent
{\em Small scalar self-coupling.} In the  limit $\lambda\to 0$
the light scalar field extends beyond the core radius $r_n$ and up
to the scale $x_\lambda\equiv \sqrt{2/\lambda}$. In this region the
gauge field can be neglected  and the first line of
Eq.~(\ref{eq::domwall}) becomes
\begin{equation}
\begin{split}
&{f''}+\lambda\left(f-f^3\right)= 0 \,.
\label{eq::kinkeq}
\end{split}
\end{equation}
It has a first integral ${f'}^2-(\lambda/2)(1-f^2)^2=0$
corresponding to the boundary value $f(\infty)=1$ and a
``half-kink'' solution
\begin{equation}
\begin{split}
&f(r_n+x)={\rm tanh}\left( x/x_\lambda\right) \,.
\label{eq::kinksol}
\end{split}
\end{equation}
Inside the boundary layer the scalar self-coupling  can be
neglected in the field equations. However, the boundary
conditions for the corresponding solution do depend on
$\lambda$. As before  matching  to the core requires
$\gamma(x)\sim \sqrt{\lambda}x$ at $x\to-\infty$ while matching
to the half-kink solution Eq.~(\ref{eq::kinksol}) gives
$f(r_n+x)\sim \sqrt{\lambda/2}x$ at $1\ll x\ll x_\lambda$. This
dependence can be eliminated by a rescaling
\begin{equation}
z =\lambda^{1/4}x\,, \quad
\gamma_\lambda( z)={\gamma(x)\over \lambda^{1/4}}\,, \quad
f_\lambda( z)={\sqrt{2}\over \lambda^{1/4}}f(r_n+x)\,.
\label{eq::newvar}
\end{equation}
In the new variables the boundary layer equations read
\begin{equation}
\begin{split}
&{  f''_\lambda} -  \gamma_\lambda^2 f_\lambda= 0 \,,\\
&  \gamma_\lambda''- f_\lambda^2 \gamma_\lambda = 0\,,
\label{eq::smallameq}
\end{split}
\end{equation}
with  the boundary conditions $\gamma_\lambda(z) \sim  z$,
$f_\lambda(z)={\cal O}(e^{- z^2/2})$ at $z\to -\infty$ and
$f_\lambda(z)\sim  z$, $\gamma_\lambda(z)={\cal O}(e^{-
z^2/2})$ at $z \to \infty$. The above system is invariant with
respect to the discrete field transformations $f_\lambda\to -
f_\lambda$, $\gamma_\lambda\to - \gamma_\lambda$, and
$f_\lambda\leftrightarrow \gamma_\lambda$, as well as the
coordinate reflection $z\to -z$.  Taking into account the
boundary conditions we find that the gauge and the scalar
fields have ``mirror'' boundary layer solutions
\begin{equation}
\gamma_\lambda(z)=-f_\lambda(-z)\,.
\label{eq::mirror}
\end{equation}
Despite the high symmery and the existence of a first integral
${\gamma_\lambda'}^2+{f_\lambda'}^2-\gamma_\lambda^2f_\lambda^2=1$
Eq.~(\ref{eq::smallameq}) cannot be integrated in quadratures,
at least not in a straightforward way. However, we can conclude that
inside the boundary layer of  width $1/\lambda^{1/4}$ the gauge
and scalar fields scale as $\lambda^{1/4}$, and hence
the boundary layer contributes  a plain ${\cal O}(1/\sqrt{n})$
correction to the vortex energy. Thus the total correction  to
the energy is dominated by the scalar cloud outside  the
boundary layer,  Eq.~(\ref{eq::kinksol}), which gives
\begin{equation}
T=2\pi\sqrt{\lambda}n\eta^2\left(1
+{4\over 3 \lambda^{1/4} \sqrt{n}}
+{\cal O}(1/\sqrt{n})\right)\,.
\label{eq::Tlamsmall}
\end{equation}
Eq.~(\ref{eq::Tlamsmall}) agrees  with a classical result for
the positive  surface tension in extreme type-I superconductors
\cite{Landau}. The scalar cloud solution itself gets the
enhanced finite-$n$ correction. It can be written as
$\delta\hspace*{-1pt}f(y)/{(\sqrt{2n}\lambda^{1/4})}$, where
$y=x/x_\lambda$ and
\begin{equation}
\delta\hspace*{-1pt}f(y)=
{9+12 y-8 e^{-2y}-e^{-4y}\over 6\sqrt{2}(2
+  e^{2y}+ e^{-2y})}\,.
\label{eq::delfres}
\end{equation}
The  details of the derivation of this result are presented in
Appendix~\ref{sec::appB}.

We should emphasize that the accuracy and convergence of the
large-$n$ expansion for noncritical vortices crucially depend
on the scale of $\lambda$, {\it cf.}
Eqs.~(\ref{eq::Tlamlarge},\hspace*{2pt}\ref{eq::Tlamsmall}).
For large scalar self-coupling the  expansion clearly breaks
down at $n={\cal O}(\sqrt\lambda)$ where the  ratio of scales
characterizing the boundary layer and the core dynamics $\delta
r_n/r_n$   is not small anymore.  At this value the dependence
on the winding number qualitatively changes. Indeed, in the
limit $\lambda\to\infty$ with fixed $n$ we should recover the
classical result \cite{Abrikosov:1956sx,Landau} where the
scalar field  core radius vanishes as $n/\sqrt{\lambda}$, a
characteristic  size of the gauge field core is ${\cal O}(1)$,
and the energy is quadratic in the magnetic flux and
logarithmic in the scalar coupling constant
\begin{equation}
T\sim  2\pi n^2\eta^2\ln(\sqrt{\lambda}/n)\,.
\label{eq::Tabrikos}
\end{equation}
For small $\lambda$ the convergence problem appears at $n={\cal
O}(1/\sqrt\lambda)$. For such $n$  the expansion in the ratio
$x_\lambda/r_n$ of the scales characterizing the dynamics of
the scalar cloud and the vortex core  breaks down, and the
structure of the solution is no longer described by the
asymptotic formulae. In particular, in the limit $\lambda\to 0$
at fixed $n$ the vortex energy is dominated by the scalar cloud
contribution, which can depend on the scalar self-coupling and
the winding number at most logarithmically.

\section{Ginzburg-Pitaevskii vortices}
\label{sec::3}
The  Ginzburg-Pitaevskii vortex  equation \cite{Ginzburg:1958}
for   a  neutral  complex scalar field   $\phi$ which describes
a Bose-Einstein condensate of weakly interacting Bose gas can
be obtained from the analysis of the previous section by
setting the gauge field to zero and rescaling the scalar
self-coupling to  $\lambda =1$. The corresponding  radial
equation reads
\begin{equation}
{1\over r}{d\over dr}\left(r{df\over dr}\right)+
 \left(1-{n^2\over r^2}-f^2\right)f  = 0\,.
\label{eq::GinPit}
\end{equation}
The structure of its solution in the limit of large $n$ is
quite distinct  from the Abrikosov-Nielsen-Olesen case
discussed above.\footnote{After a rescaling of the radial
coordinate $r\to r/\sqrt{\lambda}$ it describes the  scalar
field core of the Abrikosov-Nielsen-Olesen vortex for
$\lambda\gg n^2\gg 1$.} Now the field dynamics is essentially
different in  the vortex core $r\ll r_n$,  its tail  $r\gg
r_n$, and in  the matching region around  $r =r_n$, where
$r_n=n$ gives the leading approximation for  the size of the
region with unbroken symmetry phase. Let us consider the field
dynamics and the vortex solution in each region.

\noindent
{\em The vortex core.}
For  $r\ll r_n$ the solution $f(r)\propto r^n$ is
exponentially suppressed with $n$ so that the nonlinear cubic
term can be neglected. As $r$ approaches $r_n$ the  ${\cal
O}(1)$  linear term in the equation  becomes relevant while
the exponentially suppressed  nonlinear term can still be
omitted and we get a linear differential equation
\begin{equation}
{1\over r}{d\over dr}\left(r{df\over dr}\right)+
 \left(1-{n^2\over r^2}\right)f  = 0\,,
 \label{eq::ncoreeq}
\end{equation}
with the regular solution proportional to the $n$th Bessel
function $J_n(r)$. Since  $J^{(m)}_n(n)={\cal O}(1/n^{m+1\over
3})$,  near  $r=r_n$ all the terms of Eq.~(\ref{eq::ncoreeq})
are ${\cal O}(1/n)$ and the nonlinear part cannot be neglected.
To estimate the maximal $r$ where  Eq.~(\ref{eq::ncoreeq}) is
still valid we use an asymptotic formula
\begin{equation}
J_n(r_n+x)\sim \left(2\over n\right)^{1\over 3}
{\rm Ai}(-z)\left(1+{\cal O}(z/n^{2/3})\right)\,,
\label{eq::BesselAiry}
\end{equation}
where ${\rm Ai}(z)$ is the Airy function, $x=r-r_n$, and $z=
(2/n)^{1/3}x$. The Airy function is exponentially suppressed
for large positive values of the argument, hence  the core
approximation is valid for $r_n-r\gsim n^{1/3}$.

\noindent
{\em The vortex tail.}
In the opposite limit $r\gg r_n$ the derivative term in the
equation  is suppressed and can be treated as a perturbation.
As a result the field equation becomes algebraic. For example,
in the leading order we get
\begin{equation}
f^2-1+{n^2\over r^2}  = 0\,,
\label{eq::ntaileq}
\end{equation}
with the solution
\begin{equation}
f(r)=\left(1-{n^2\over r^2}\right)^{1\over 2}\,.
 \label{eq::ntailsol}
\end{equation}
The higher order terms  can easily be obtained by iterations to
any desired order  in $1/r$. The first three terms read
\begin{eqnarray}
f(r) &=& \left(1-{n^2\over r^2}\right)^{1\over 2}
-{n^2\over r^4}\left(1-{n^2\over 2r^2}\right)
\left(1-{n^2\over r^2}\right)^{-{5\over 2}}
\nonumber\\
&+&{n^2\over r^6}\left(8+{3n^2\over 2r^2}
-{n^4\over 2r^4}+{n^6\over 8r^6}\right)
\left(1-{n^2\over r^2}\right)^{-{11\over 2}}
+{\cal O}(1/r^8)\,.
\label{eq::nfseriesr}
\end{eqnarray}
The expansion Eq.~(\ref{eq::nfseriesr}) formally breaks down
for  $x= {\cal O}(n^{1/3})$. Indeed, let us consider the
behavior of the above series at
$x={\kappa}^{1-\alpha}n^\alpha/2$ for some power $\alpha$. Here
the factor ${\kappa}^{1-\alpha}/2$ with a constant $\kappa$ is
introduced for convenience. Then the series
Eq.(\ref{eq::nfseriesr}) can be written as an expansion
\begin{equation}
f(r)=\left({\kappa \over n}\right)^{1-\alpha\over 2}
\sum_{m=1}^\infty f_m \rho^m\,,
\label{eq::nfseriesrho}
\end{equation}
where  $\rho={\kappa ^{3\alpha-3}/ n^{3\alpha-1}}$. Each $f_m$
can in turn be expanded in  $\tau=(\kappa/n)^{1-\alpha}$. For
the first three terms we get
\begin{eqnarray}
&&f_1=1-{3\over 8}\tau+{23\over 128}\tau^2+\ldots\,,
\nonumber\\
&&f_2={1\over 2}+{7\over 16}\tau-{117\over 256}\tau^2+\ldots\,,
\label{eq::nfseriesrtau}\\
&&f_3={73\over 8}+{601\over 64}\tau +{1215\over 1024}\tau^2+\ldots\,.
\nonumber
\end{eqnarray}
For $\alpha\le 1/3$  the expansion parameter $\rho$ is not
suppressed at large $n$ and the series
Eq.~(\ref{eq::nfseriesr}) is not defined (though for the
boundary case $\alpha=1/3$ we have  $\rho=1/\kappa^2$ and the
series may converge for sufficiently large $\kappa$). At the
same time the expansion parameter $\tau$ is not suppressed  for
$\alpha\ge 1$. Thus the allowed interval for the expansion in
negative  powers of $n$ is   $1/3<\alpha<1$. For $\alpha=1/2$
both expansions in $\rho$ and $\tau$ convert  into a series
in $1/\sqrt{n}$. Note that by choosing a smaller or larger value
of $\alpha$ we boost the convergence of the expansion  in $\tau$
or $\rho$, respectively, and for a given $\alpha$ the parameter
$\kappa$ can be tuned to further improve  the convergence of
the series.

\noindent
{\em The matching region.}
The problem now is to derive a large-$n$ expansion in the
matching region $-n^{1/3}\lsim x \lsim n^{\alpha}$. First we
note that in this interval the solution  is suppressed  at
least as $1/n^{5\alpha-1\over 2}$, {\it i.e.}  at $n\to\infty$
the boundary layer does not develop. Moreover,  the core
solution at  $x={\cal O}(1)$ and the tail solution at $x ={\cal
O}(n^{1/3})$ have the property  $f^{(m)}(r)  ={\cal
O}(1/n^{m+1\over 3})$, which means that they can be
polynomially matched over the corresponding  interval. A linear
matching of the leading core and tail solutions gives a simple
analytic formula
\cite{Penin:2020cxj}
\begin{equation}
f(r)=\left\{
\begin{array}{l}
 C'J_n(r)\,,  \quad r\le r_n\,,  \\[3pt]
 (\delta/ 2n)^{1/2}
 \left(1+{(r-r_n)/\delta}\right),\,
 r_n< r< r_n+\delta\,,  \\[3pt]
 \left(1-{n^2/  r^2}\right)^{1/2}\,,
 \quad r_n+\delta\le r\,,
\end{array}
\right.
\label{eq::linmat}
\end{equation}
where $C'={3^{1/4}\pi^{1/2}\over 2^{1/2}}$,
$\delta/n^{1/3}={2^{2/3}\pi \over 3^{5/6}\Gamma^2(2/3)}$, and
$\Gamma(z)$ is the Euler gamma-function. In the limit
$n\to\infty$ the function defined by Eq.~(\ref{eq::linmat})
becomes continuous and smooth. As has been discussed above, the
higher orders of the perturbative expansion of the core
solution in the nonlinear term at  $x= {\cal O}(1)$ as well as
the perturbative expansion of the tail  solution in the
derivative term at  $x= {\cal O}(n^{1/3})$ are not suppressed
by   extra powers of $n$ (though both expansion may {\it de
facto} converge). Thus  Eq.~(\ref{eq::linmat}) gives the
asymptotic solution  up to  ${\cal O}(1/n^{1/3})$ corrections.
It can be used to obtain  the analytic result for the vortex
energy at $n\to\infty$. The total energy consists of the
logarithmically divergent rotational part $T_{rot}$ and the
nonrotational finite part $T$ \cite{Ginzburg:1958}. The
rotational contribution determined by the angular part of the
kinetic term reads
\begin{equation}
T_{rot}=\eta^2\int{|\partial_\theta\phi({\bfm r})/r|^2
}{\rm d}^2{\bfm r}
\approx 2\pi n^2\eta^2\left(\ln\left({R/ r_n}\right)
-{1\over 2}\right)\,,
\label{eq::Trot}
\end{equation}
where $R\gg r_n$ is the radial integration cutoff (the physical
radius of rotating condensate). For the remaining part
saturated by the field potential we get $T=\pi  n^2 \eta^2$,
where a half of the contribution comes from the vortex tail. As
expected  the logarithmic term in Eq.~(\ref{eq::Trot}) agrees
with Eq.~(\ref{eq::Tabrikos})  upon identification of $R$ with
the characteristic size  of the gauge field core and $r_n$ with
the one of the scalar core.

\begin{figure}[t]
\begin{center}
\includegraphics[width=8cm]{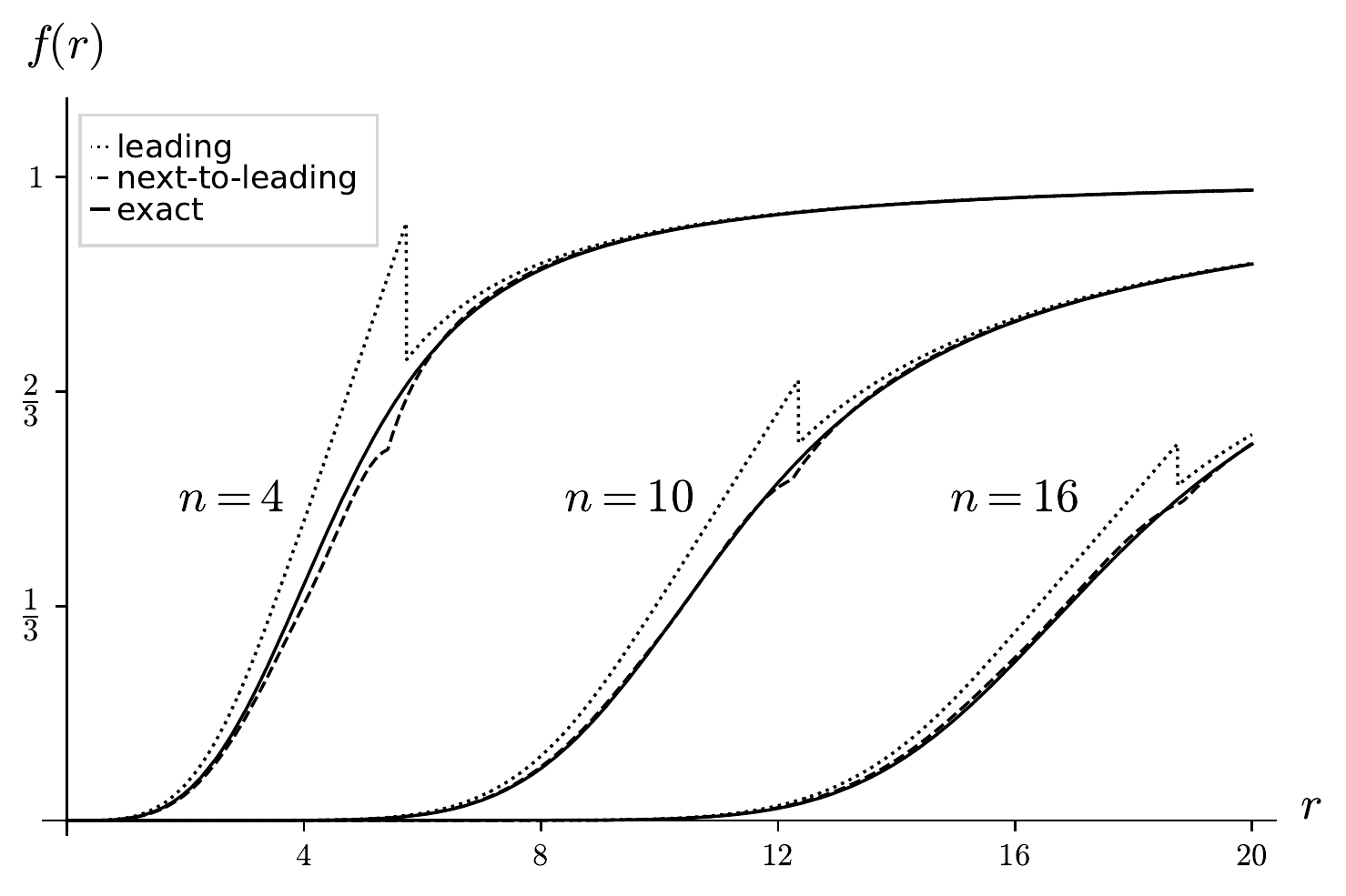}
\end{center}
\caption{\label{fig::7}   The  numerical  solution of the exact
vortex equations for a neutral scalar field $f(r)$ (solid line),
the leading asymptotic solution Eq.~(\ref{eq::linmat})  (dotted
line), and the next-to-leading approximation  (dashed line) for
different $n$.}
\end{figure}

To obtain the  solution at ${\cal O}(1/n^{1/3})$  and beyond,
the coefficients of the vortex equation  in the matching region
are expanded in $x/r_n$. Defining
\begin{equation}
f(r) ={2^{5/6}\over n^{1/3}}\,\chi\left(-z\right)\,,
\label{eq::chidef}
\end{equation}
we get the leading order $n$-independent  nonlinear equation
for the function $\chi(z)$
\begin{equation}
{d^2\chi\over dz^2}-z\chi-2\chi^3=0\,.
\label{eq::PanleveII}
\end{equation}
Eq.~(\ref{eq::PanleveII}) is  a special case of the
Painlev\'e~II transcendent. It is known to have a solution
$\chi_0(z)$ with the following asymptotic behavior at
$z\to\infty$
\cite{Hastings:1980}
\begin{equation}
\begin{split}
&\chi_0(-z)\sim \left({z\over 2}\right)^{1\over 2}\,,\\
&\chi_0(z)\sim {\rm Ai}(z)\,.
\label{eq::chiasym}
\end{split}
\end{equation}
The corresponding function  $f(r)$ matches the core solution
$\sqrt{2}J_n(r)$ at  negative $x= {\cal O}(n^{1/3})$ and the
leading order tail solution $f_1$ at  positive $x= {\cal
O}(\sqrt{n})$. In the next-to-leading order the equation for
$\chi(z)$ becomes
\begin{equation}
{d^2\chi\over dz^2}-{1\over 2^{1/3}n^{2/3}}{d\chi\over dz}
-z\left(1+{3\over 2^{4/3}}{z\over n^{2/3}}\right)\chi-2\chi^3=0\,.
\label{eq::nchieq}
\end{equation}
The relevant  solution must exponentially decay at $z\to\infty$
and match  the expansion of the next-to-leading tail solution
$f(r_n+x)$ at $x= \sqrt{\kappa n}/2$ through ${\cal O}(1/n)$.
It can only be found numerically. Note that since the expansion
parameter $x/n$  varies from $1/n$ to $1/\sqrt{n}$ through the
matching region,  the convergence of the series there is not
uniform. In general the expansion of the tail solution in
$\rho$ matches the asymptotic expansion of $\chi_0(z)$ at large
negative $z$ while the series in $\tau$ for each $f_m$ is
connected with the corrections to the equation for $\chi(z)$.

The results of numerical analysis for the  neutral scalar field
vortices with $n=4,\,10,\,16$  are presented in
Fig.~\ref{fig::7}.  For  the leading order approximation the
linear matching is used  and in the  next-to-leading order  the
numerical solution of Eq.~(\ref{eq::nchieq}) is matched to the
first two terms of Eq.~(\ref{eq::nfseriesr}) at
$r_n+\sqrt{n/2}$, which corresponds to $\kappa = 2$.

\section{Summary}
\label{sec::sum}

We have elaborated a method of  expansion in  inverse powers of
a topological quantum number.  The method is quite general and
can be applied to the study of topological solitons in a theory
where the corresponding  quantum number is associated with a
ratio of dynamical scales. When applied to axially symmetric
vortices with large winding number $n$ the expansion is in
negative noninteger powers of $n$ and in general is not
uniform. In the large-$n$ limit the complex nonlinear vortex
dynamics unravels. In particular, for the
Abrikosov-Nielsen-Olesen vortices at critical coupling and in
the limit of large or small scalar self-coupling the field
equations become integrable. At the same time for the
Ginzburg-Pitaevskii vortices in  a weakly interacting
Bose-Einstein condensate the field equation  reduces to an
algebraic one. The method  yields simple asymptotic formulae
for the shape and parameters capturing the main features of the
{\it giant} vortices. The accuracy of the asymptotic result can
be systematically improved by calculating the finite-$n$
corrections. For near-critical vortices the approximation works
remarkably well all the way down to very low $n$ after
including the leading ${\cal O}(1/\sqrt{n})$ corrections.  In
the case of extremely  small or large scalar self-coupling
$\lambda$ the large-$n$ expansion  converges and gives a
reliable approximation  for all winding numbers satisfying
the condition $n\gg \sqrt{\lambda},~1/\sqrt{\lambda}$.

\vspace{2mm}
\noindent
{\bf Acknowledgement}\\[1mm]
A.P.  is grateful to Valery Rubakov  for a discussion of
noncritical vortices. The work of  A.P. was supported in part
by NSERC and the Perimeter Institute for Theoretical Physics.
The work of  Q.W. was supported through the NSERC USRA program.

\appendix

\section{Calculation of the critical vortex parameters}
\label{sec::appA}
{\em Calculation   of ${w_\infty}$.}
The integral in Eq.~(\ref{eq::boundsol}) is logarithmically
divergent at $w(x)\to 0$ and can be decomposed  as follows
\begin{eqnarray}
\int_{w_0}^{w(x)}\frac{{\rm d}w}{(e^{2w}-2w-1)^{1/2}}
&=&-\int_{w_0}^{w(x)}\frac{{\rm d}w}{\sqrt{2}w}+\int_{0}^{w(x)}
\left(\frac{1}{(e^{2w}-2w-1)^{1/2}}+\frac{1}{\sqrt{2}w}\right){\rm d}w
\nonumber\\
&+&\int_{w_0}^{0}\left(\frac{1}{(e^{2w}-2w-1)^{1/2}}
+\frac{1}{\sqrt{2}w}\right){\rm d}w\,,
\label{eq::wintdec}
\end{eqnarray}
where the first  term is divergent, the second term is ${\cal
O}(w(x))$, and the last term is a constant which we denote as
$\ln({w_\infty/w_0})/\sqrt{2}$. Thus Eq.~(\ref{eq::boundsol})
at  $w(x)\to 0$ becomes
\begin{equation}
x=-{1\over\sqrt{2}}\ln\left({w(x)\over w_\infty}\right)
+{\cal O}(w(x))\,,
\label{eq::asyminfty}
\end{equation}
which after exponentiation gives the second line of
Eq.~(\ref{eq::wasym}).

\noindent
{\em Calculation  of $B$ and $C$.}
By changing the integration variable $w'{\rm d}y={\rm d}w$
the integral in Eq.~(\ref{eq::delwres}) can be transformed
as follows
\begin{eqnarray}
\int_{z}^{\infty}w'^2(y){\rm d}y&=&
-\left(-{w(z)\over 2}\right)^{1\over 2}
\left(1+{4\over 3}w(z)\right)
\nonumber\\
&+&\int_{w(z)}^{0}\left[{\left(e^{2 w}-2w-1\right)^{1\over 2}}
-\left(-{2w}\right)^{1\over 2}
\left(1+{1\over 4w}\right)\right]{\rm d}w\,.
\label{eq::wp2int}
\end{eqnarray}
At  $z\to-\infty$ up to ${\cal O}(1/z)$ corrections the first
term is just $-z^3/3$ and the second term is a constant which
we denote as $B$. This gives
\begin{equation}
\int_{z}^{\infty}{w'^2(y)\over w'^2(z)}{\rm d}y\sim
-\frac{z}{3}+{B\over z^2}+\ldots\,,
\label{eq::Cintasym}
\end{equation}
where the ellipsis stands for the exponentially suppressed terms.
Then the  integral in Eq.~(\ref{eq::delwres})
can be decomposed  as follows
\begin{eqnarray}
\int_{0}^{x}\int_{z}^{\infty}\frac{w'^2(y)}{w'^2(z)}
{\rm d}y{\rm d}z&=&-\frac{x^2}{6}+\int_{-\infty}^{x}
\left(\frac{z}{3}+\int_{z}^{\infty}
\frac{w'^2(y)}{w'^2(z)}{\rm d}y\right){\rm d}z
\nonumber \\
&-&\int^{0}_{-\infty}
\left(\frac{z}{3}+\int_{z}^{\infty}\frac{w'^2(y)}{w'^2(z)}
{\rm d}y\right){\rm d}z\,,
\label{eq::Cintdec}
\end{eqnarray}
where the second term vanishes at $x\to-\infty$ as $-B/x$  and
the last term is a constant which has to be cancelled by $C$ in
Eq.~(\ref{eq::delwres}) to provide the correct asymptotic
behavior, Eq.~(\ref{eq::delwasym}).

\noindent
{\em Calculation  of $D$.}
The asymptotic behavior of the integral in
Eq.~(\ref{eq::delwres}) at $x\to\infty$  can be computed via
the following decomposition
\begin{eqnarray}
\int_{0}^{x}\int_{z}^{\infty}\frac{w'^2(y)}{w'^2(z)}{\rm d}y\,{\rm d}z
&=&\frac{x}{2\sqrt{2}}-\int^{\infty}_{x}\left(-\frac{1}{2\sqrt{2}}
+\int_{z}^{\infty}\frac{w'^2(y)}{w'^2(z)}{\rm d}y\right){\rm d}z
\nonumber\\
&+&\int_{0}^{\infty}\left(-\frac{1}{2\sqrt{2}}
+\int_{z}^{\infty}\frac{w'^2(y)}{w'^2(z)}{\rm d}y\right){\rm d}z\,,
\label{eq::nuintdec}
\end{eqnarray}
where the second term vanishes at $x\to \infty$ and the last
term is a constant. By substituting Eq.~(\ref{eq::nuintdec})  into
Eq.~(\ref{eq::delwres}) at  $x\to \infty$ we obtain the second
line of Eq.~(\ref{eq::delwasym}).

\section{Finite-$\mbox{\bfm n}$ corrections to the
noncritical vortex solution}
\label{sec::appB}
{\em  Large  scalar self-coupling.}
By expanding  Eq.~(\ref{eq::fieldeqfa}) about $r_n+\delta r_n$
we get
\begin{equation}
\begin{split}
&\delta\hspace*{-1pt}f= (1-\gamma_\lambda^2)^{-{1\over 2}}
\gamma_\lambda\left[(x-1)\gamma_\lambda
-\delta\gamma_\lambda\right]\,,\\
& \delta\gamma_\lambda''
-2\left(1-3\gamma_\lambda^2\right)
\delta\gamma_\lambda =
\gamma_\lambda'+4(x-1)\gamma_\lambda^3\,,
\label{eq::delgeq}
\end{split}
\end{equation}
which defines  an inhomogeneous  differential equation on the
function $\delta\gamma_\lambda(y)$, where $y=x-x_0$. The
boundary conditions for this equation are
$\delta\gamma_\lambda(\infty)=0$ and
$\delta\gamma_\lambda(-x_0)=1/2$, where the latter is
determined by matching to the core solution for the gauge field
at $r_n+\delta r_n$. Note that $\delta\hspace*{-1pt}f$  and
$\delta\gamma_\lambda''$ are not continuous  at this point due
to the singular character of the matching region in the limit
$\lambda\to\infty$ discussed previously. Due to the  existence
of the first integral for Eq.~(\ref{eq::largelameq}) the
homogeneous part of the solution $\delta\gamma^h_\lambda$
satisfies the first-order equation
\begin{equation}
{d\over d\gamma_\lambda}
\left[\gamma_\lambda^2\left(2-\gamma_\lambda^2\right)
{d\delta\gamma^h_\lambda\over d\gamma_\lambda}
-2\gamma_\lambda\left(1-\gamma_\lambda^2\right)
\delta\gamma^h_\lambda\right]= 0\,.
\label{eq::delgeq2}
\end{equation}
One of its solutions is given by the translational zero mode
$\gamma_\lambda'(y)=2\,{\rm sech}(\sqrt{2}y)\, {\rm
tanh}(\sqrt{2}y)$. The second solution reads
\begin{equation}
\delta\gamma^h_\lambda(y)=
{{\rm sech}(\sqrt{2}y)\, {\rm tanh}(\sqrt{2}y)\over 8\sqrt{2}}
\left(6\sqrt{2}y-4\,{\rm coth}(\sqrt{2}y)
+{\rm sinh}(2\sqrt{2}y)\right)\,.
\label{eq::delgh}
\end{equation}
With the solutions of the homogeneous equation at hand, the
solution of the full equation
is straightforward and gives Eq.~(\ref{eq::delgres}).

\noindent
{\em Small scalar self-coupling.}
The expansion of Eq.~(\ref{eq::fieldeqfa}) with $a(r)=1$ about
$r_n+\delta r_n$  yields
\begin{equation}
\delta\hspace*{-1pt}f'' +2(1-3f^2)\delta\hspace*{-1pt}f= -\sqrt{2}f'\,,
\label{eq::delfeq1}
\end{equation}
with the boundary conditions  $\delta f(\infty)=0$ and $\delta
f(0)=0$. The condition at $y=0$ is determined by matching to
the boundary layer solution at $x={\cal O}(1/\lambda^{1/4})$.
The homogeneous part of the solution satisfies the  first-order
equation
\begin{equation}
{d\over d\hspace*{-1pt}f}
\left[{ \left(1-f^2\right)^2\over 2}
{d\delta\hspace*{-1pt}f^h\over d\hspace*{-1pt}f}
+f\left(1-f^2\right)
\delta\hspace*{-1pt}f^h\right]= 0\,.
\label{eq::delfeq2}
\end{equation}
It has the translational zero mode $f'(y)={\rm
sech}^2(y)$ and the  second solution
\begin{equation}
\delta\hspace*{-1pt}f^h(y)=
{{\rm sech}^2(y)\over 16}
\left(12y+8\,{\rm sinh}(2y)
+\,{\rm sinh}(4y)\right)\,,
\label{eq::delfh}
\end{equation}
which after straightforward calculation gives
Eq.~(\ref{eq::delfres}).

\end{document}